\documentclass[11pt]{article}
\usepackage{moriond,epsfig}

\def\Journal#1#2#3#4{{#1} {\bf #2}, #3 (#4)}

\def\PRD{{\em Phys. Rev.} D}

\def\be{\begin{equation}}
\def\ee{\end{equation}}
\def\bea{\begin{eqnarray}}
\def\eea{\end{eqnarray}}

\def\ltap{\ \raisebox{-.4ex}{\rlap{$\sim$}} \raisebox{.4ex}{$<$}\ }
\def\gtap{\ \raisebox{-.4ex}{\rlap{$\sim$}} \raisebox{.4ex}{$>$}\ }
\newcommand{\deltaatm}{\mbox{$\Delta  m^2_{\mathrm{atm}} \ $}}
\newcommand{\deltasol}{\mbox{$ \Delta  m^2_{\odot} \ $}}

\newcommand{\betabeta}{\mbox{$(\beta \beta)_{0 \nu}  $}}
\newcommand{\hbeta}{$\mbox{}^3 {\rm H}$ $\beta$-decay \ }

\newcommand{\meff}{\mbox{$\left|  < \! m  \! > \right| \ $}}
\newcommand{\eV}{\mbox{$ \  \mathrm{eV} \ $}}
\newcommand{\deltatre}{\mbox{$ \ \Delta m^2_{32} \ $}}
\newcommand{\deltadue}{\mbox{$ \ \Delta m^2_{21} \ $}}
\newcommand{\ueuno}{\mbox{$ \ |U_{\mathrm{e} 1}|^2 \ $}}

\newcommand{\uetre}{\mbox{$ \ |U_{\mathrm{e} 3}|^2  \ $}}
\begin{document}
\vspace*{4cm}
\title{Neutrino Mass Spectra, CP-Violation 
and Neutrinoless Double-Beta Decay\footnote{Presented by S. Pascoli}}

\author{S. M. Bilenky$^{(i,j)}$
\footnote{Also at: Joint Institute for Nuclear Research, Dubna, Russia}
,~S. Pascoli$^{(j,l)}$~and~
S. T. Petcov$^{(j,l)}$ 
\footnote{Also at: Institute of Nuclear Research and
Nuclear Energy, Bulgarian Academy of Science, Bulgaria}
}

\address{
$^{(i)}$ Physik Department, Technische Universitat Munchen,
D-85748, Garching, Germany \\
$^{(j)}$ Scuola Internazionale Superiore di Studi Avanzati, 
I-34014 Trieste, Italy\\
 $^{(l)}$ Istituto Nazionale di Fisica Nucleare, 
Sezione di Trieste, I-34014 Trieste, Italy\\}

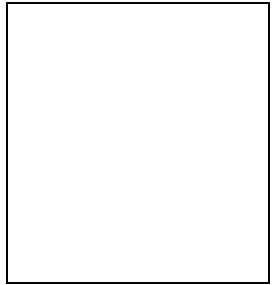
\maketitle\abstracts{
Assuming 3-$\nu$ mixing and  massive Majorana neutrinos,
we present some  implications of the  
oscillation solutions of the solar 
and atmospheric neutrino problems
and of the results of the CHOOZ experiment
for the predictions of the
effective Majorana mass in
neutrinoless double beta-decay. 
If the present or upcoming \betabeta-decay 
searches give a positive result, the Majorana nature
of massive neutrinos will be established.
From the determination of the value of the
effective Majorana mass, it would be possible 
 to obtain information 
on the neutrino mass spectrum.
With additional information 
on the absolute value of neutrino masses,
it might be possible to infer 
if  CP-parity is violated in the leptonic sector.
}

\section{Introduction}

 With the accumulation of more and stronger 
evidences for oscillations of the
atmospheric\cite{SKatm00}
and solar\cite{SKYSuz00} neutrinos,  
caused by  neutrino mixing 
(see, e.g., \cite{BiPe87}), 
the problem of the nature of massive neutrinos 
emerges as one of the fundamental problems 
in the studies of neutrino mixing. 
Massive neutrinos can be Dirac or Majorana particles.
In the former case they possess 
a conserved lepton charge and
distinctive antiparticles, 
while in the latter there is no conserved 
lepton charge 
and massive neutrinos  are truly neutral 
particles identical with their antiparticles
(see, e.g., \cite{BiPe87}).
Thus, the question of the nature of 
massive neutrinos is directly related 
to the question of the 
basic symmetries of the 
fundamental particle interactions.

The present and upcoming experiments 
devoted to study neutrino oscillations
will allow to make a big step forward
in  understanding  the 
patterns of neutrino mass squared differences and 
of $\nu$-mixing but won't be able to determine 
the absolute value of $\nu$-masses and 
to answer the question regarding 
the nature of massive neutrinos.

 The  \hbeta experiments, studying the electron spectrum, 
are sensitive to the electron (anti-)neutrino mass  
$m_{\nu_e}$ and can give information  on 
 the absolute value of neutrino masses:
 the Troitzk~\cite{MoscowH3} and Mainz~\cite{Mainz} 
 experiments present bounds read
  $ m_{\nu_e}  <  2.5 
\eV$~\cite{MoscowH3} and 
$  m_{\nu_e} <  2.9  \eV$~\cite{Mainz} (at 95\%~C.L.)
and there are prospects to 
increase the sensitivity  
of the \hbeta experiments
to  $m_{\nu_e} \sim (0.3 - 1.0)$ eV by 
the  KATRIN \cite{KATRIN} experiment.

The problem of the nature of massive neutrinos
can be addressed in experiments 
in which   the total lepton charge $L$
is not conserved and changes by two units,
$\Delta L = 2$.
The process most sensitive to the 
existence of massive Majorana neutrinos 
(coupled to the electron)
is the neutrinoless double $\beta$ 
($\betabeta-$) decay of
certain even-even nuclei (see, e.g., \cite{BiPe87}): 
$(A,Z) \rightarrow (A,Z+2) + e^{-} + e^{-}$.
If the $\betabeta-$ decay is generated 
only by the 
left-handed  charged current weak interaction
through the
exchange of virtual light
massive Majorana neutrinos, 
the probability amplitude of this process 
is proportional 
to the ``effective Majorana mass parameter'':
\begin{equation}
\meff \equiv  | U_{\mathrm{e} 1}^2~ m_1 +
U_{\mathrm{e} 2}^2~ m_2 + U_{\mathrm{e} 3}^2~ m_3 |,
\label{meff}
\end{equation}

\noindent where $m_j$ is the mass of the 
Majorana neutrino $\nu_j$  
and $U_{\mathrm{e}j} = |U_{\mathrm{e}j}| e^{i \alpha_j /2}$ 
is the element of the neutrino (lepton) 
mixing matrix, with $\alpha_j$ the CP-violating phase. 
\meff depends only on two Majorana
CP-violating phase differences\cite{BHP80} 
$\alpha_{j1} \equiv \alpha_{j}- \alpha_1$, $j=2,3$. 
If  CP-parity is conserved we have\cite{LW81,BNP84}
 $\alpha_{j1} =k \pi$, $k=0,1,2...$.
  Many experiments are searching  for 
$\betabeta-$decay. 
No  indications that this process takes place were found so far.
A rather  stringent constraint on the 
value of \meff 
was obtained in the $^{76}$Ge Heidelberg-Moscow 
experiment\cite{76Ge00} 
$ \meff < (0.35 \div 1.05) \ \mathrm{eV}$ 
 and in the  IGEX one\cite{IGEX00}
$\meff < (0.33 \div 1.35) \ \mathrm{eV}$ (both at $90\%$C.L.).
  Higher sensitivity to the value of 
$\meff$ is planned to be 
reached in several $\betabeta-$decay experiments
of a new generation\cite{NEMO3}. 
The NEMO3, 
scheduled to start in 2001, and the planned CUORE
experiments will have 
a sensitivity to values of $\meff \cong 0.1~$eV.
A sensitivity to $\meff \cong 10^{-2}~$eV,
is planned to be achieved 
in the GENIUS and  EXO experiments.


\section{Predictions for the Effective
Majorana Mass Parameter.}

Assuming 3-$\nu$ mixing 
and massive Majorana neutrinos,
 we present some  implications of the neutrino 
oscillation fits of the solar 
and atmospheric neutrino data
and of the results of the 
reactor long baseline CHOOZ and Palo Verde 
experiments\cite{SKYSuz00,SKatm00,Fogli00,Gonza3nu,CHOOZ},
for the predictions of the
effective Majorana mass parameter 
\meff, which controls 
the \betabeta-decay rate. 
The present article represents a brief review of the detailed results obtained in Ref.\cite{BPP1}.
Earlier studies on the subject include, e.g.,
Refs.\cite{bbpapers}.
The case of 4-$\nu$ mixing 
has been treated also in detail  in  Ref.\cite{BPP2}.
The predicted values of  \meff
depend strongly on the type of the
neutrino mass spectrum, on the solution of 
the solar neutrino problem, as well as 
on the values of the two
Majorana CP-violating phases,
present in the lepton mixing matrix.
We consider here three types of 
mass spectra: hierarchical ($m_1 \ll m_2 \ll m_3$),
with inverted hierarchy ($m_1 \ll m_2 \sim  m_3$) ,
and with three quasi-degenerate  neutrinos
 ($m_1 \sim m_2 \sim  m_3$).

If the neutrino mass 
spectrum is hierarchical, only the contributions
to \meff due to the exchange of the two heavier 
Majorana neutrinos $\nu_{2,3}$ can be relevant
and \meff depends only on 
one CP-violating phase, $\alpha_{31}-\alpha_{21}$.
The effective mass can be given in terms
of the oscillating parameters \deltasol,\deltaatm and 
$|U_{\mathrm{e}3}|^2$, which is constrained by the CHOOZ data, as:
\be
\meff \simeq \left| \sqrt{\deltasol} (1 - |U_{\mathrm{e}3}|^2) 
\sin^2 \theta_\odot + 
\sqrt{\deltaatm} |U_{\mathrm{e}3}|^2 e^{i(\alpha_{31} - \alpha_{21})} \right|.
\label{eqmasshierarchy01} 
\ee
For the SMA and LOW-QVO solutions 
one has $\meff \ltap 4.0\times 10^{-3}~{\rm eV}$.
For the LMA solution we get
$\meff \ltap 8.5-10 \times 10^{-3}~{\rm eV}$
if one uses the results of the analyzes in
\cite{SKYSuz00,Fogli00} and \cite{Gonza3nu}.
The maximal values of \meff correspond to
CP-conservation and 
$\nu_{2}$ and $\nu_{3}$ having 
identical CP-parities, $\phi_2 = \phi_3$.
For all three solutions there are no 
significant lower bounds on \meff because
mutual compensations
between the terms contributing
to \meff, corresponding to the exchange of different 
virtual massive Majorana neutrinos, are possible.

   In the case of the inverted hierarchy
mass spectrum, 
the dominant contribution to \meff
is due to 
the two heavier Majorana neutrinos 
$\nu_{2,3}$ and \meff is determined 
by \deltaatm, $\sin^22\theta_{\odot}$,
the CP-violating phase $\alpha_{31}-\alpha_{21}$
and by $|U_{\mathrm{e}1}|^2$,
 constrained by the CHOOZ data as:
\be
\meff \simeq \sqrt{\deltaatm} (1 - \ueuno) \sqrt{1 - \sin^22\theta_{\odot} 
\sin^2\left( \alpha_{31} - \alpha_{21} \right)/2}.
\label{meffinvmh3}
\end{equation}
The effective Majorana mass 
can be considerably larger
than in the case of a 
hierarchical neutrino mass spectrum:
$\meff \ltap (6.8 - 8.9)\times 10^{-2}~{\rm eV}$. 
The maximal \meff corresponds to
CP-conservation and 
$\phi_2 = \phi_3$ ($\alpha_{31} -\alpha_{21}  =0$); 
it is possible for all three
solutions of the $\nu_{\odot}-$problem.
A lower bound of $\meff \gtap (3 - 4) \times 10^{-2}~{\rm eV}$
is present in the case of the SMA solution,
while for the LMA and LOW-QVO 
solutions values of 
$\meff \ll  10^{-2}~{\rm eV}$
are possible. Both for the
LMA and LOW-QVO 
solutions there exist 
relatively large ``just-CP-violation'' region
(a value of \meff in this
region would unambiguously signal the existence of
CP-violation in the lepton sector, caused by 
Majorana CP-violating phases)
 of \meff.
A measurement of \meff,
\deltaatm, $\sin^22\theta_{\odot}$,
$(1 - |U_{\mathrm{e}1}|^2)$ can allow to get 
direct information on the CP-violation in the lepton sector
and on the value of
the CP-violating phase $\alpha_{31} -\alpha_{21}$.

   For  quasi-degenerate neutrinos,
 we have $\meff \sim m$, 
where $m$ is the common neutrino mass
 constrained by the
$^{3}$H $\beta-$decay experiments.
\meff
depends also on $\theta_{\odot}$,
$|U_{\mathrm{e} 3}|^2$ (constrained by the CHOOZ data), 
and on two physical CP-violating phases,
$\alpha_{21}$ and $\alpha_{31}$:
\be
\meff \simeq m \left| \cos^2 \theta_\odot (1- \uetre) +  
\sin^2    \theta_\odot (1- \uetre)   
 e^{i \alpha_{21}} +|U_{\mathrm{e} 3}|^2 e^{i \alpha_{31}} \right|.
\label{meffmdeg}
\end{equation}

The maximal value of \meff is determined by
the value of $m$, $\meff \leq m$ and
 is limited by the upper bounds 
obtained in the $^{3}$H $\beta-$decay
\cite{MoscowH3,Mainz}
and in the \betabeta-decay 
\cite{76Ge00,IGEX00} 
experiments: 
$\meff < (0.33 - 1.05)~{\rm eV}$.
The existence of a significant lower bound 
on \meff for the LMA and the LOW-QVO solutions
depends on the ${\rm min}|\cos2\theta_{\odot}|$.
The latter varies with the analysis:
using the results of Ref.\cite{SKYSuz00} 
one finds $\meff \gtap (0.1 - 0.2)~m$
~~($\meff \gtap (0.2 - 0.3)~m$) for the
LMA (LOW-QVO) solution.
According to the  results 
of the analysis\cite{Fogli00,Gonza3nu}
one can have $\cos2\theta_{\odot} = 0$
and therefore there is no 
significant lower bound
on \meff for both solutions. 
There exist  
``just-CP-violation'' regions for the
LMA and LOW-QVO solutions, in which
\meff can be in the range of sensitivity of
the future \betabeta-decay experiments, 
while in the case of the SMA MSW solution
there is no such region.
The knowledge  of \meff, $m$,
$\theta_{\odot}$ and $|U_{\mathrm{e}3}|^2$ 
would imply a non-trivial
constraint on the two CP-violating phases
$\alpha_{21}$ and $\alpha_{31}$.

\section{Conclusions}

If \betabeta-decay will be 
detected  by present or upcoming experiments,
we will conclude that neutrinos are massive
Majorana particles and that the total 
lepton charge $L$ is not conserved.
  The observation of the
\betabeta-decay with a rate
corresponding to 
$\meff \gtap 0.02~$eV,
which is in the range of sensitivity of the 
future \betabeta-decay experiments, 
can provide unique information 
on the neutrino mass spectrum. 
A measured value of 
$\meff \gtap (2 - 3)\times 10^{-2}~$eV
would strongly disfavor (if not rule out), 
(under the general assumptions of 
3-neutrino mixing, \betabeta-decay generated 
only by the charged (V-A) current weak interaction 
via the exchange of the three Majorana neutrinos,
neutrino oscillation solutions of the solar 
neutrino problem and atmospheric
neutrino anomaly) the possibility of a 
hierarchical neutrino mass spectrum,
while a value of 
$\meff \gtap (2 - 3)\times 10^{-1}~$eV
would rule out the 
hierarchical neutrino mass spectrum,
strongly disfavor the 
inverted  hierarchy one  and favor
the quasi-degenerate spectrum.

From the determination of the value of \meff
constraints on the absolute value of 
neutrino masses could be inferred.
Having additional information on the value of 
neutrino masses or the type of neutrino mass spectrum, 
we could obtain also information 
on the CP-violation in the lepton sector,
and - if CP-invariance holds - on the 
relative CP-parities
of the massive Majorana neutrinos.
 For the LMA MSW and LOW solutions of the solar-$\nu$ anomaly,
it would be possible to find  either an allowed range 
of values for $m_1$ ( if $\deltadue = \deltasol$ 
or $\deltatre = \deltasol$ and $m_1 > 10^{-2} \eV$) 
or an upper bound on  $m_1$ 
(if $\deltatre = \deltasol$ and $m_1 < 10^{-2} \eV$).
With additional information on $m_1$
we could establish if CP is violated in the lepton sector
 and either the values of the CP-violating phases 
or the type of CP-parity patterns of the neutrinos.
 For the SMA MSW solution of the solar-$\nu$ anomaly,
we could obtain the value of $m_1$ (if $\deltadue = \deltasol$ 
or $\deltatre = \deltasol$ and $m_1 > 10^{-2} \eV$) 
or an upper bound on  $m_1$ 
(if $\deltatre = \deltasol$ and $m_1 < 10^{-2} \eV$).
In this case we  have  no information on 
the CP-violation in the lepton sector.

\section*{Acknowledgments}
Si. P. would like to thank the Organizer Committee for the pleasant
and stimulating atmosphere of this conference, 
held in such a beautiful location.

\end{document}